%% file: zhh_tianjp.tex
\documentclass[article,tightenlines,amsmath,amssymb]{revtex4}
\usepackage{graphicx}
\usepackage{dcolumn}
\usepackage{bm}
\usepackage{psfig}
\usepackage{rotating}
\usepackage{enumerate}
\usepackage{subfigure}

\usepackage[usenames]{color}
\usepackage{xspace}
\usepackage{hepparticles}
\begin{document}
\normalsize
\parskip=5pt plus 1pt minus 1pt

\title{Study of Higgs Self-coupling at ILC}

\include{author}

\date{\today}

\begin{abstract}

In this Analysis we investigated the possibility of the measurement of Higgs self-coupling at ILC through the process $e^++e^-\rightarrow ZHH$ using fast simulation data. So far two combinations of decay modes: $Z\rightarrow q \bar q,H\rightarrow b\bar b, H\rightarrow WW^*$ and $Z\rightarrow l \bar l, H\rightarrow b\bar b, H\rightarrow b\bar b$ were studied. Our preliminary results show that it is very challenging to suppress the huge standard model backgrounds effectively. 

\end{abstract}


\maketitle

\section{Introduction}

It is well accepted that the discovery of a Higgs-like boson is not enough to fully understand the mechanism of electro-weak symmetry breaking (EWSB) and mass generation. The Higgs self-coupling can be a non-trivial probe of the Higgs potential and probably the most decisive test of the EWSB mechanism. In the standard model framework, the Higgs potential $V(\Phi)=\lambda(\Phi^2-\frac{1}{2}v^2)^2$, where $\Phi$ is an isodoublet scalar field and $v\approx 246$ GeV is the vacuum expectation value of its neutral component, is uniquely determined by the self-coupling $\lambda$. Obviously, determination of the Higgs mass, which satisfies $m_H^2=2\lambda v^2$ at tree level, can provide an indirect information about the self-coupling. The measurement of the trilinear self-coupling $\lambda_{HHH}=6\lambda v$ offers direct independent determination of the Higgs potential shape, which is the topic of this analysis.

\begin{center}
\begin{figure}[htbp]
\subfigure{
\label{triiinear:a}
\includegraphics[width=6cm,height=4cm]{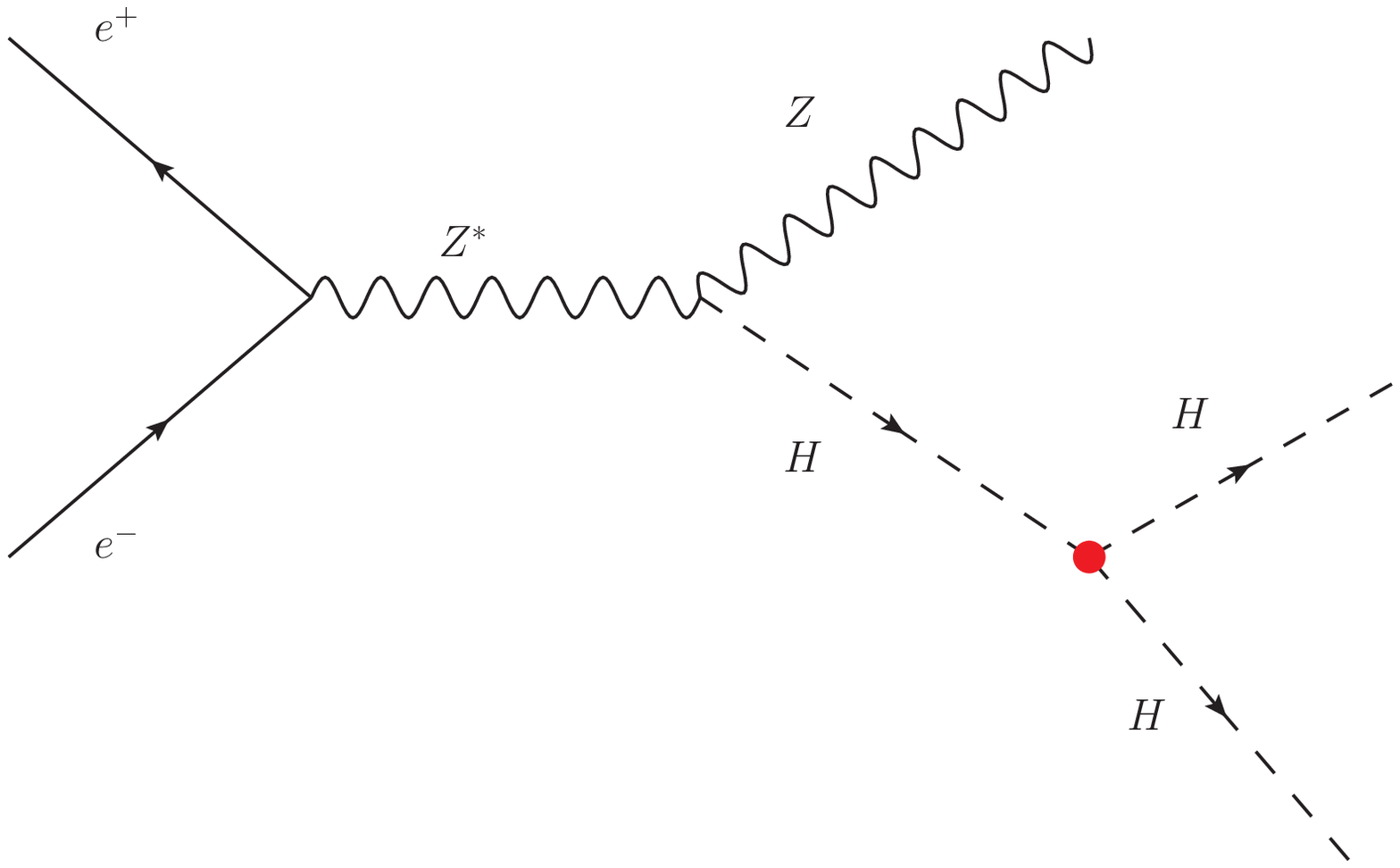}
}
\subfigure{
\label{trilinear:b}
\includegraphics[width=6cm,height=4cm]{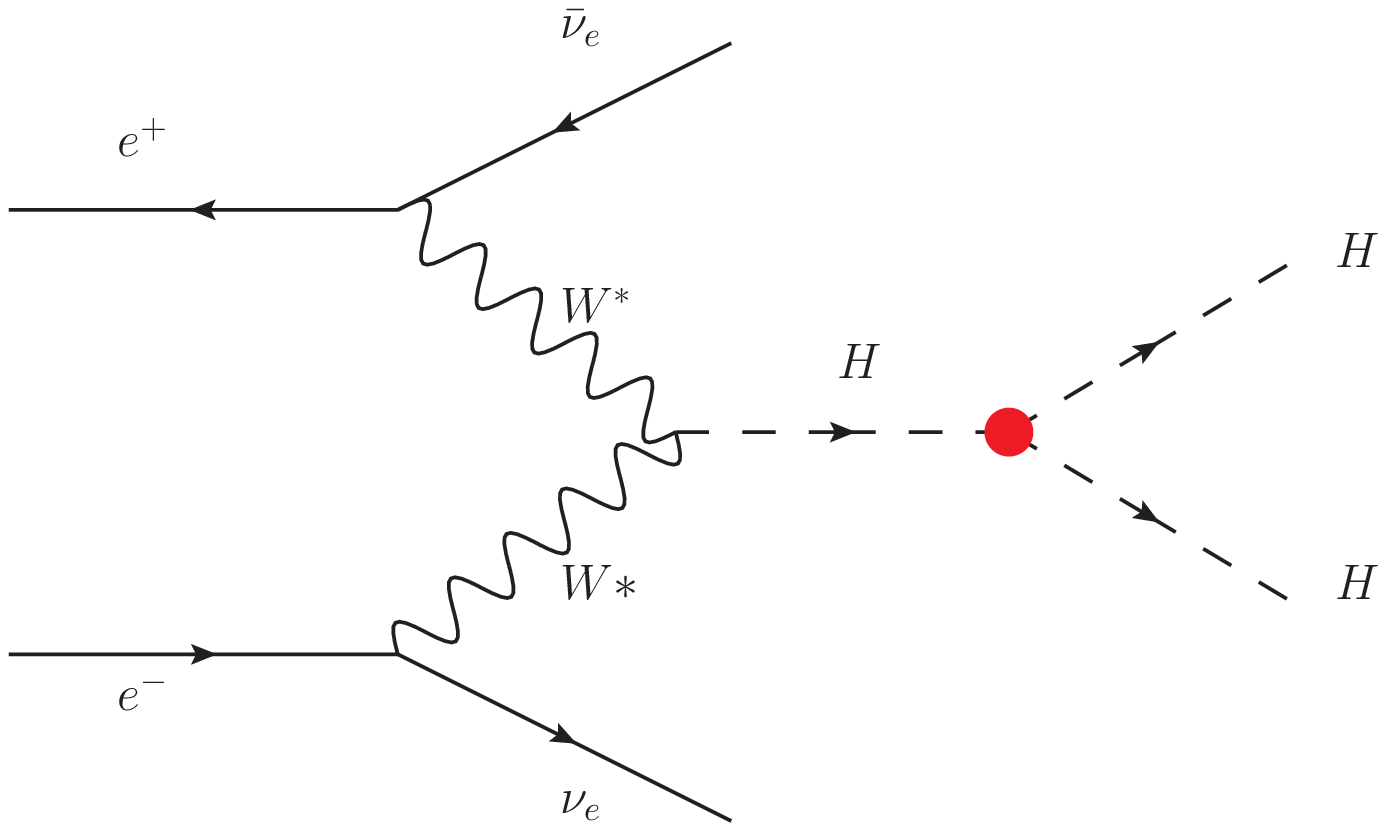}
}
\caption{Leading processes involving trilinear Higgs self-coupling: (Left) Double Higgs-strahlung; 
(Right) WW fusion.}
\label{trilinear}
\end{figure}
\end{center}

The trilinear Higgs self-coupling can be measured at ILC through the two leading processes: double Higgs-strahlung ~\cite{Hstrahlung1,Hstrahlung2, Hstrahlung3} and WW fusion ~\cite{Hstrahlung2, Hstrahlung3,wwfusion1,wwfusion2,wwfusion3,wwfusion4}, which are shown in Fig.\ref{trilinear}. The former is expected to dominate around the center of mass energy of 500 GeV and the latter to take it over at higher energy. In this analysis we focus on the double Higgs-strahlung process $e^++e^-\rightarrow ZHH$ for the Higgs mass of $M_H=120$ GeV and the center of mass energy of $\sqrt{s}=500$ GeV with the integrated luminosity 2 ${\rm ab}^{-1}$. 

Depending on the different decay modes of $Z$ and $H$, there are different methods to identify the signal events. Table I shows several most promising combinations of decay modes for $e^++e^-\rightarrow ZHH$ and their branching ratios. Modes 1 and 3 are studied in Ref. ~\cite{takubo}. We study the other two modes in this analysis.

\begin{center}
\begin{table}[htpb]
\caption{Most promising modes for $e^++e^-\rightarrow ZHH$}
\end{table}
\begin{tabular}{c|c|c|c|c}
\hline \hline Decay Mode & $Z\rightarrow$ & $H_1\rightarrow$ & $H_2\rightarrow$ & Branching Ratio \\
\hline 1 & $q\bar q$ & $b\bar b$ & $b\bar b$ & 34\% \\
\hline 2 & $q\bar q$ & $b\bar b$ & $WW^*   $ & 14\% \\
\hline 3 & $\nu\bar \nu$ & $b\bar b$ & $b\bar b$ & 9.8\% \\
\hline 4 & $l\bar l$ & $b\bar b$ & $b\bar b$ & 4.9\% \\
\hline
\end{tabular}
\end{center}

\section{Simulation}
The simulations of signal events ($e^++e^-\rightarrow ZHH$) and possible background events ($e^++e^-\rightarrow t\bar t, ZZZ, W^+W^-Z, ZZ, ZH$) were done by Physsim ~\cite{physsim}. In Physsim the helicity amplitudes are calculated by the HELAS library ~\cite{helas}. The phase space integration and the four momenta generation are performed by BASES/SPRING ~\cite{bases}. Parton showering and hadronization are carried out by PYTHIA6.4 ~\cite{pythia}, where final-state $\tau$ leptons are decayed by TAUOLA ~\cite{tauola} in order to handle their polarizations correctly. The detector simulation was done by JSFQuickSimulator, which implements the GLD geometry and other detector-performance related parameters ~\cite{jsfqksm}. 

It is worth mention of that the simulations were performed without the beam polarization but with the initial-state radiation, beam width and beamstrahlung. Then the cross sections used here are shown in Table II. An integrated luminosity of 2 ${\rm ab}^{-1}$ is assumed in this analysis.
\begin{center}
\begin{table}[htpb]
\caption{Cross sections of the related processes}
\end{table}
\begin{tabular}{c|c|c|c|c|c|c}
\hline \hline Process & $e^++e^-\rightarrow ZHH$ & $e^++e^-\rightarrow t\bar t$ & $e^++e^-\rightarrow ZZZ$ & $e^++e^-\rightarrow W^+W^-Z$ & $e^++e^-\rightarrow ZZ$ & $e^++e^-\rightarrow ZH$\\
\hline Cross section & 152 ab & 530 fb & 800 ab & 36 fb & 515 fb & 70 fb \\
\hline
\end{tabular}
\end{center}

\section{Analysis}
\subsection{$e^++e^-\rightarrow ZHH\rightarrow (q\bar q)(b\bar b)(WW^*)$}
\begin{center}
\begin{figure}[htbp]
\centering
\includegraphics[width=7.5cm,height=7.5cm]{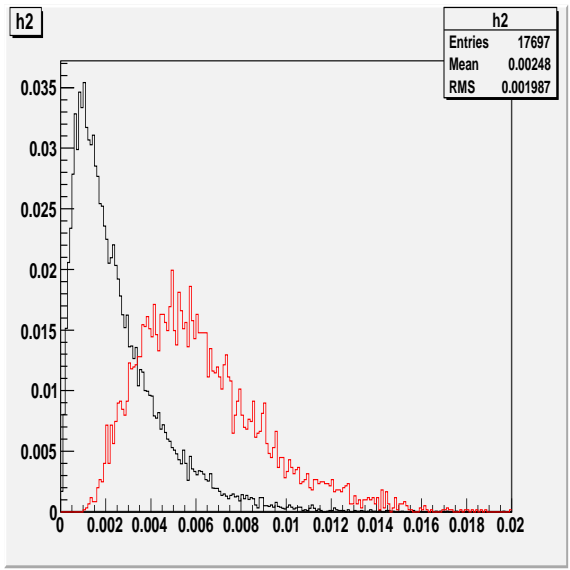}
\caption{Distribution of $Y_{cut}$, where black is for signal and red is for $t\bar t$ background.}
\label{ycut}
\end{figure}
\end{center}

The full hadronic decays of $W$ and $W^*$ were investigated. In this mode the final state of a candidate signal event contains of 8 jets, two of which are $b$ jets. To select the signal events, first we find all the good tracks and require the number of tracks be greater than 20. We then try to combine tracks with a small $Y$ value to a current jet cluster, where the $Y$ value between two momenta $p_1,p_2$ is defined as $Y(p_1,p_2)=\frac{M^2(p_1,p_2)}{E_{vis}}$, with $M(p_1,p_2)$ being the invariant mass of $p_1, p_2$ and $E_{vis}$ the total visible energy. We continue the jet clustering until there are 7 jets left, because the two jets coming from $W^*$ are very close to each other which means the Y-value between them is very small, thereby being likely to be clustered as one jet. At this point we calculate the $Y$ values for all the pairs from these 7 jets and choose  the minimum denoted by $Y_{cut}$. The $Y_{cut}$ distributions of signal events and background events (here we consider the $t\bar t$ events as background) are shown in Fig.\ref{ycut}. The 7 jets are combined by minimizing the $\chi^2$ which is defined as $$\chi^2=\frac{(M(b,\bar b)-M_H)^2}{\sigma^2_{H_1}}+\frac{(M(W,W^*)-M_H)^2}{\sigma^2_{H_2}}+\frac{(M(q,\bar q)-M_Z)^2}{\sigma^2_{Z}}+\frac{(M(q,\bar q')-M_W)^2}{\sigma^2_{W}}$$ where $M(q,q')$ is the reconstructed invariant mass of jet $q$ and jet $q'$, $M_H$, $M_Z$ and $M_W$ are the mass of $H$, $Z$ and $W$, respectively, and $\sigma_{H_1},\sigma_{H_2},\sigma_{Z} and \sigma_{W}$ are their corresponding mass resolutions.

In order to further suppress the background, we require that $\chi^2<20, 90{\rm GeV}<M(H_1)<130{\rm GeV}, 110{\rm GeV}<M(H_2)<150{\rm GeV}, 70{\rm GeV}<M(Z)<110{\rm GeV}, Y_{cut}>0.0076$, where the asymmetry of two Higgs mass requirement is due to their different decay modes. The preliminary result of this cut-based analysis is shown in Table III. Though we can still add other cuts like $b$ tagging requirement, the signal events will become too few to be observed. It seems very challenging to reject the huge $t\bar t$ background in this mode.

We are going to investigate the semi-lepton decays of $W$ and $W^*$.

\begin{center}
\begin{table}[htpb]
\caption{Cut statistics of $e^++e^-\rightarrow ZHH\rightarrow (q\bar q)(b\bar b)(WW^*)$}
\end{table}
\begin{tabular}{c|c|c}
\hline \hline Process & $ZHH\rightarrow (q\bar q)(b\bar b)(WW^*) $ & $t\bar t$ \\
\hline theoretical & 18.3 & 1062000 \\
\hline pre-selection & 12.6 & 483949 \\
\hline $\chi^2<20$ & 5.2 & 65144 \\
\hline $90GeV<M_{H_1}<130GeV$ & 5.1 & 63157 \\
\hline $110GeV<M_{H_2}<150GeV$ & 3.6 & 36670 \\
\hline $90GeV<M_Z<110GeV$ & 3.5 & 34359 \\
\hline $Y_{cut}>0.005$ & 2.3 & 8454 \\
\hline $Y_{cut}>0.0076$ & 1.1 & 2644 \\
\hline
\end{tabular}
\end{center}

\subsection{$e^++e^-\rightarrow ZHH\rightarrow (l\bar l)(b\bar b)(b\bar b)$}
In this mode a candidate signal event contains two leptons and four $b$ jets, where we only consider the $Z$ boson decaying into $e^+e^-$ and $\mu^+\mu^-$. The two isolated charged lepton tracks are required to have an energy greater than 20 GeV and the energy deposited in the cone of $20^\circ$ around each lepton track be less than  20 GeV. We then force the other tracks to four jets and combine the four jets by minimizing the $\chi^2$ defined by $$\chi^2=\frac{(M(b,\bar b)-M_H)^2}{\sigma^2_{H_1}}+\frac{(M(b,\bar b)-M_H)^2}{\sigma^2_{H_2}}+\frac{(M(l,\bar l)-M_Z)^2}{\sigma^2_{Z}}.$$ 

\begin{center}
\begin{figure}[htbp]
\centering
\includegraphics[width=7.5cm,height=7.5cm]{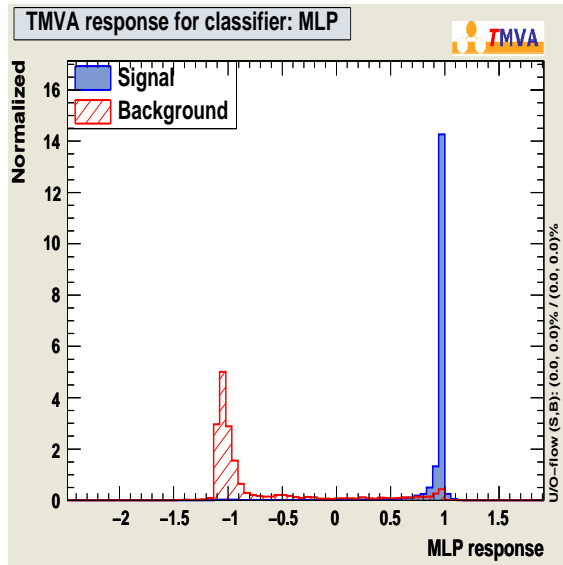}
\caption{The classifier obtained by neural net training for signal and $t\bar t$ background.}
\label{mvatt}
\end{figure}
\end{center}

Table IV shows that 15.4 signal events survived the pre-selection but with thousands times more background events left.  In order to reject the background effectively, while keeping a reasonable signal  efficiency, we used the neural net method MLP in the TMVA package ~\cite{tmva} which gives some useful classifiers. Here we mainly consider the five kinds of backgrounds that are shown in Table IV. First we separately do the neural net analysis between the signal and each of the five kinds of backgrounds. For each event we can get five classifiers to separate signal and backgrounds. Figure \ref{mvatt} histograms the classifiers obtained by the MLP method for the signal and $t\bar t$ samples. 

We then add some more cuts on the five classifiers denoted by $mva\_tt, mva\_zzz, mva\_wwz, mva\_zz$ and $mva\_zh$ as shown in Table IV. We further impose cuts on the $Z$ mass, $Y_{cut}$, and require $b$ tagging, which is based on the number of tracks with 2.5$\sigma$ separation from the interaction point. Our preliminary result is listed in Table IV. The final cut is applied with the neural net for the signal and the $ZZZ$ background after all of the above cuts. We end up with 3 signal events with 0.82 $ZZZ$ events left, while four backgrounds are eliminated. The result shows that the ZHH events can be observed in this mode with the significance $\frac{S}{\sqrt{S+B}} \sim 1.5\sigma$.

\begin{center}
\begin{table}[htpb]
\caption{Cut statistics of $e^++e^-\rightarrow ZHH\rightarrow (l\bar l)(b\bar b)(b\bar b)$}
\end{table}
\begin{tabular}{c|c|c|c|c|c|c}
\hline \hline Process & $ZHH$ & $t\bar t$ & $ZZZ$ & WWZ & ZZ & ZH \\
\hline generated &1M & 4.5M & 500K & 750K & 1.25M & 250K\\ 
\hline theoretical & 304 & 1062000 & 1600 & 72300 & 1030000 & 140000\\
\hline pre-selection & 15.4 & 9023 & 125 & 1943 & 3560 & 1618\\
\hline 
$mva\_tt>0.98$     &    &&&&&\\
$mva\_wwz>1.0$  &   &&&&&\\
$mva\_zz>0.97$     &  11.7 & 312 & 12.9 & 12.7 & 16.5 & 5.6\\
$mva\_zh>0.97$    &   &&&&&\\
$mva\_zzz>0$         &   &&&&&\\
\hline $70GeV<M_Z<110GeV$ & 9.7 & 106 & 11.7 & 7.5 & 16.5 & 0.56 \\
\hline $Y_{cut}>0.015$ & 9.1 & 91.3 & 10.6 & 6.9 & 6.6 & 0\\
\hline $2b(H_1) (N_{off}>0)$ & 6.3 & 28 & 5.5 & 1.8 & 0 & 0\\
\hline $2b(H_2) (N_{off}>1)$ & 3.5 & 0.71 & 2.3 & 0 & 0 & 0\\
\hline $mva\_zzz>0.86$ & 3.0 & 0 & 0.82 & 0 & 0 & 0\\
\hline
\end{tabular}
\end{center}

\section{Summary}
The two modes, $e^++e^-\rightarrow ZHH\rightarrow (q\bar q)(b\bar b)(WW^*)$ and $e^++e^-\rightarrow ZHH\rightarrow (l\bar l)(b\bar b)(b\bar b)$, were investigated for the purpose of the measurement of the trilinear Higgs self-coupling at ILC for $M_H=120$ GeV, $\sqrt{s}=500$ GeV and the integrated luminosity of 2 ${\rm ab}^{-1}$. The former mode is very difficult to use for the signal observation, while the latter mode can be useful to observe the self-coupling.
\acknowledgments
We would like to thanks all the members of the ILC physics subgroup ~\cite{ilcphy} for useful discussions. This study is supported in part by KEK, Center of High Energy Physics, Tsinghua University and the JSPS Core University Program.

\end{document}

%% file: author.tex
\author{
Junping Tian$^{1}$, Keisuke Fujii$^{2}$, Yuanning Gao$^{1}$
\\
\vspace{0.2cm}
{\it
$^{1}$ Tsinghua University, Beijing 100084, People's Republic of China\\
$^{2}$ High Energy Accelerator Research Organization (KEK), Tsukuba, Japan }
}